\documentclass[conference, final]{IEEEtran}
\IEEEoverridecommandlockouts
\usepackage{hyperref}       
\usepackage{url}            
\usepackage{booktabs}       
\usepackage{amsfonts}       
\usepackage{nicefrac}       
\usepackage{microtype}      
\usepackage{xcolor}         
\usepackage{multirow}
\usepackage{listings}
\usepackage{pgfplots}
\usepgfplotslibrary{dateplot}
\pgfplotsset{compat=1.17}
\usepackage{subcaption}
\usepackage{caption}
\usepackage{hyperref}
\usepackage{enumitem}
\usepackage{algorithm}
\usepackage{hyperref}
\usepackage{graphicx}
\usepackage{tikz-cd}
\usepackage{tcolorbox}
 \usepackage[frozencache,cachedir=mint]{minted}
 \usepackage{fancyvrb}
\usepackage{pgfplots}
\usepackage{xurl}
\usepackage[noadjust]{cite}
\usepackage{graphicx}
\usepackage{algpseudocode}
\usepackage{algorithm}
\usepackage{tabularx}
\usepackage{balance}
\usepackage{makecell}
\usepackage{tikz}
\usepackage{pgfplots}
\pgfplotsset{compat=1.17}
\pgfdeclarelayer{background}
\pgfsetlayers{main}
\usepackage{multirow}
\usetikzlibrary{external}
\usepackage{rotating}
\usepackage{amsmath}

\usepackage{algorithm, algpseudocode}
\usepackage{setspace}
\usepackage{xcolor}
\usepackage{gensymb}
\usepackage{placeins}
\usepackage{enumitem}
\usetikzlibrary{decorations.pathreplacing,positioning}
\usepackage[noadjust]{cite}
\usepackage{dirtytalk}

\usepgfplotslibrary{groupplots}

\pgfplotsset{compat=1.17}
 \usepackage{etoolbox}
\makeatletter
\patchcmd{\@verbatim}
  {\verbatim@font}
  {\verbatim@font\small}
  {}{}
\makeatother

\hypersetup{
   breaklinks=true,   
   colorlinks=true,   
}

\definecolor{archtBlue}{RGB}{0, 114, 178}
\definecolor{lightBlue}{RGB}{0, 114, 240}
\definecolor{archtRed}{RGB}{213, 94, 0}
\definecolor{lightRed}{RGB}{250, 94, 0}
\definecolor{archtGreen}{RGB}{0, 128, 0}
\definecolor{lightGreen}{RGB}{0, 230, 0}
\definecolor{archtPurple}{RGB}{128, 0, 128}
\definecolor{archtOrange}{RGB}{230, 159, 0}
\definecolor{archtPink}{RGB}{204, 121, 167}

\newcommand{\hpcorpus}[0]{\textsc{HPCorpus}}

\newcommand{\comp}[0]{\textsc{MonoCoder}}

\newcommand{\mypara}[1]{\textit{#1}}

\makeatletter
\newcommand*{\rom}[1]{\expandafter\@slowromancap\romannumeral #1@}
\makeatother

\def\BibTeX{{\rm B\kern-.05em{\sc i\kern-.025em b}\kern-.08em
    T\kern-.1667em\lower.7ex\hbox{E}\kern-.125emX}}

\DeclareRobustCommand*{\IEEEauthorrefmark}[1]{%
  \raisebox{0pt}[0pt][0pt]{\textsuperscript{\footnotesize\ensuremath{#1}}}}
  \usepackage[misc]{ifsym}

\ifCLASSINFOpdf

\else

\fi

\DeclareRobustCommand*{\IEEEauthorrefmark}[1]{%
  \raisebox{0pt}[0pt][0pt]{\textsuperscript{\footnotesize\ensuremath{#1}}}}
  \usepackage[misc]{ifsym}

\begin{document}

\title{\comp{}: Domain-Specific Code Language Model for HPC Codes and Tasks\\
\thanks{This research was supported by the Israeli Council for Higher Education (CHE) via the Data Science Research Center, Ben-Gurion University of the Negev, Israel; Intel Corporation (oneAPI CoE program); Pazy Foundation; and the Lynn and William Frankel Center for Computer Science. Computational support was provided by \textcolor{blue}{\href{https://console.breckenridge.cloud/}{HPE HPC \& AI Cloud}}, \textcolor{blue}{\href{https://www.intel.com/content/www/us/en/developer/tools/devcloud/overview.html}{Intel Developer Cloud}}, and \textcolor{blue}{\href{https://platform.openai.com/tokenizer}{the NegevHPC project}}. Part of this work was completed when Niranjan Hasabnis was at Intel Labs and Gal Oren with NRCN.}
}
\author{\IEEEauthorblockN{Tal Kadosh\IEEEauthorrefmark{1,2},
Niranjan Hasabnis\IEEEauthorrefmark{3},
Vy A. Vo\IEEEauthorrefmark{4},
Nadav Schneider\IEEEauthorrefmark{1,2},
Neva Krien\IEEEauthorrefmark{},
Mihai Capotă\IEEEauthorrefmark{4},\\
Abdul Wasay\IEEEauthorrefmark{4},
Guy Tamir\IEEEauthorrefmark{5},
Ted Willke\IEEEauthorrefmark{4},
Nesreen Ahmed\IEEEauthorrefmark{4},
Yuval Pinter\IEEEauthorrefmark{1},
Timothy Mattson\IEEEauthorrefmark{} and
Gal Oren\IEEEauthorrefmark{6,7}}\\
\IEEEauthorblockA{\IEEEauthorrefmark{1}Ben-Gurion University, 
\IEEEauthorrefmark{2}IAEC,
\IEEEauthorrefmark{3}Code Metal,
\IEEEauthorrefmark{4}Intel Labs, 
\IEEEauthorrefmark{5}Intel, 
\IEEEauthorrefmark{6}Technion, 
\IEEEauthorrefmark{7}Stanford University}

{\tt\small talkad@post.bgu.ac.il, niranjan@codemetal.ai, vy.vo@intel.com, }\\
{\tt\small nadavsch@post.bgu.ac.il, nevo.krien@gmail.com, mihai.capota@intel.com, }\\
{\tt\small abdul.wasay@intel.com, guy.tamir@intel.com, ted.willke@intel.com,}\\
{\tt\small nesreen.k.ahmed@intel.com, pintery@bgu.ac.il, tim@timmattson.com, }\\
{\tt\small galoren@stanford.edu}}

\maketitle

\begin{abstract}
With easier access to powerful compute resources, there is a growing trend in AI for software development to develop large language models (LLMs) to address a variety of programming tasks. Even LLMs applied to tasks from the high-performance computing (HPC) domain are huge in size and demand expensive compute resources for training. This is partly because LLMs for HPC tasks are obtained by finetuning existing LLMs that support several natural and/or programming languages. We found this design choice confusing --- \emph{why do we need LLMs trained on natural languages and programming languages unrelated to HPC for HPC-specific tasks?}

In this line of work, we aim to question choices made by existing LLMs by developing smaller language models (LMs) for specific domains --- we call them \emph{domain-specific LMs}. Specifically, we start with HPC as a domain and build an HPC-specific LM, named \comp{}, which is orders of magnitude smaller than existing LMs but delivers better performance on non-HPC and HPC codes. Specifically, we pre-trained \comp{} on an HPC-specific dataset (named \hpcorpus{}) of C and C++ programs mined from GitHub. We evaluated the performance of \comp{} against state-of-the-art multi-lingual LLMs. Results demonstrate that \comp{}, although much smaller than existing LMs, outperforms other LLMs on normalized-perplexity tests (in relation to model size) while also delivering competing CodeBLEU scores for high-performance and parallel code generations. In other words, results suggest that \comp{} \textit{understands} HPC code better than state-of-the-art LLMs. 

\comp{} source code is available at our GitHub
\textcolor{blue}{\href{https://github.com/Scientific-Computing-Lab/MonoCoder}{repository}}.



\end{abstract}

\section{Introduction}

Recent breakthroughs in the field of AI have led significant attention to language models (LMs) due to their advanced capabilities in natural language processing (NLP)~\cite{min2021recent}. Large language models (LLMs), particularly exemplified by models such as GPT-3~\cite{floridi2020gpt} and its successors~\cite{bubeck2023sparks}, deployed in the conversational form of ChatGPT~\cite{openai2023chatgpt}, have demonstrated the potential to grasp intricate linguistic structure and semantics, sparking exploration of their applicability beyond NLP.

In parallel, the field of high-performance computing (HPC) has been tackling increasingly complex and data-intensive problems~\cite{reed2022reinventing}. The field of HPC has experienced advancements in hardware, software, and algorithms, resulting in improvements in computational performance and efficiency~\cite{dongarra2022hpc, reed2023hpc}. Combining the two trends, integrating LMs into HPC workflows has emerged as a compelling avenue for innovation~\cite{chen2023lm4hpc}. For instance, several recent efforts have explored the application of LLMs in assisting HPC programmers in automatically inserting OpenMP pragmas or MPI functions in code~\cite{chen2023learning, harel2023learning, kadosh2023pragformer, kadosh2023advising, nichols2023modeling, schneider2023mpi, shen2023multigraph, chen2024position}, overcoming the limitations of existing static tools~\cite{harel2020source, milewicz2021negative, mosseri2020compar, prema2017identifying, prema2019study}.

Although existing LLMs have shown remarkable results on programming tasks~\cite{hou2023large}, such as code generation, or bug fixing, we found several limitations. First, we found that they perform surprisingly poorly on the HPC-related programming tasks, such as code parallelization and vectorization~\cite{kadosh2023pragformer, kadosh2023advising, nichols2023modeling, chen2023lm4hpc, nicholsCanLargeLanguage2024}. Our observations are corroborated by the latest work by Nichols et al.~\cite{nicholsCanLargeLanguage2024} which specifically evaluates parallel code generation performance of start-of-the-art LLMs, such as GPT-4 by building a benchmark named ParEval (Parallel Code Generation Evaluation). Nichols et al. find that \say{\textit{LLMs are significantly worse at generating parallel code than they are at generating serial code}}. Moreover, they further comment that \say{\textit{the poor performance of LLMs on ParEval benchmark indicates that further efforts are necessary to improve the ability of LLMs to model parallel code and/or create new LLMs that are specialized for parallel code generation.}}

This finding led us to ponder several research questions: \textit{(i) How well do existing LLMs perform on domain-specific tasks such as those for the HPC domain}, \textit{(ii) Will more domain-specific training data help in improving the performance?}, \textit{(iii) When applying LMs for a specific domain, do we finetune existing LMs, or should we train them from scratch?} This last question is specifically important in the context of the enormous training costs of LLMs, which is the second limitation of existing LLMs. As an example, HPC-Coder~\cite{nichols2023modeling}, a recently-introduced LM for HPC tasks, is obtained by finetuning PolyCoder~\cite{xu2022systematic} on an HPC dataset; PolyCoder itself is a code LM (not specific to HPC) that is trained on a corpus made up of 249 GB of programs written in 12 programming languages. As another example, HPC-GPT~\cite{ding2023hpc}, LM4HPC~\cite{chen2023lm4hpc}, and AutoParLLM~\cite{mahmud2023autoparllm} rely on LLaMA as their base model --- with up to 34 billion parameters --- or even on GPT-4. Such setups looked counter-intuitive to us --- \emph{Is it not enough to train an LM on HPC-specific languages only? In other words, why do we need an LM trained on Java or Python --- with tens or hundreds of billion parameters --- for HPC-specific tasks?} More importantly, we believe that such \emph{domain-specific LMs} would be computationally as well as financially efficient to train.

In this paper, we hypothesize that HPC-specific LMs (e.g., smaller LMs that are designed and trained specifically on HPC datasets) would perform better than existing LMs for HPC tasks. We perform two experiments to validate this hypothesis. The purpose of the first experiment is to build a smaller LM that performs similar, if not better, than existing LLMs in terms of generic language understanding tasks (such as code completion). Towards that end, in the first experiment, we build a smaller LM, called \comp{}, by reducing the number of layers of PolyCoder by a factor of 4 and pretraining \textit{only} on C and C++ codes. We then empirically validate that \comp{}, despite being a smaller model than PolyCoder, achieves a comparable perplexity score to PolyCoder in generic code completion task. The purpose of the second experiment is then to evaluate the performance of \comp{} on HPC-specific tasks. Towards that end, we obtain the CodeBLEU scores of parallel code generation models with increasing context.


We exploit one insight that we learned while building \comp{}. Specifically, we found that existing code LMs capture ``local'' semantics of code structures that leads to their degraded performance on the code completion task. An example of ``local'' semantics could be a variable named \texttt{i} is an index variable of a \texttt{for} loop (in most of the programs for that matter). We address this limitation in \comp{} by implementing a code pre-processing scheme that eliminates any local semantics that the LLM may capture. Our experimental evaluation demonstrates that \comp{} outperforms existing code LMs in virtually all of the code completion settings and context lengths, with and without applying local semantics elimination (henceforth, LSE). Moreover, and in contrast to \comp{}, the performance of other LMs degraded considerably when LSE was applied, suggesting their reliance on local semantics.

\textbf{Contributions.} This paper makes following contributions:
\begin{itemize}
    \item By drawing insights from the limitations of existing LLMs, we propose a domain-specific, small language model, called \comp{}, for tasks related to high-performance computing.
    \item We design a pre-processing method, called Local Semantics Elimination (LSE), that eliminates syntactical constructs that could lead to local semantics.
    \item We compare \comp{} against PolyCoder and GPT-3.5 for general-purpose programming tasks as well as tasks related to HPC programming.
    \item Finally, we also measure the parallel code generation performance of \comp{} and other LLMs over a dataset of 20k OpenMP codes by calculating the CodeBLEU score. Our results demonstrate that \comp{}, although orders of magnitude smaller in size, performs similar to these LLMs in language comprehension, while outperforming them in HPC-specific tasks. 
\end{itemize}

The remainder of the paper is organized as follows: 
In \autoref{hpcor} we delve into the details of the HPC code dataset \hpcorpus{}.
Then \autoref{CompCode} provides an in-depth exploration of \comp{}, our language model pre-trained on \hpcorpus{}.
In \autoref{tokomp}, we introduce LSE, our code preprocessing step designed for HPC code. 
In \autoref{section:intrinsic_eval}, we perform evaluations of \comp{} and state-of-the-art models on non-HPC and HPC tasks.
Finally, we conclude the paper and outline directions for the future research in \autoref{conclusion}.

\begin{table}[!t]
\centering
\begin{tabular}{c||r|r|r|r}
\hline
                 & \textbf{Repos} & \textbf{Size (GB)} & \textbf{Files (\#)} & \textbf{Functions (\#)} \\ \hline
C       & 144,522        & 46.23             & 4,552,736           & 87,817,591              \\ \hline
C++     & 150,481        & 26.16             & 4,735,196           & 68,233,984              \\ \hline
\end{tabular}
\vspace{0.2cm}

\caption{Statistics on the subset of \hpcorpus{} dataset~\cite{kadosh2023quantifying} that we used in our study: $\sim$300k repos, $\sim$70 GB, $\sim$9M files, and $\sim$155M functions across C and C++ code from GitHub.
}
\label{tab:hpcorpus}
\end{table}

\begin{figure*}[!htbp]
    \centering
\noindent
\begin{minipage}{0.25\linewidth}
\begin{lstlisting}[breaklines=true, breakatwhitespace=true, basicstyle=\footnotesize\ttfamily, columns=fullflexible]
// Source code:
int main() {
   int r[2800 + 1];
}
\end{lstlisting}
\end{minipage}%
\begin{tikzpicture}
  \draw[->] (0,0) -- (0.4,0);
\end{tikzpicture}%
\hspace{0.4cm}
\begin{minipage}{0.35\linewidth}
\begin{lstlisting}[breaklines=true, breakatwhitespace=true, basicstyle=\footnotesize\ttfamily, columns=fullflexible]
// LSE:
int func_252() {
  int arr_88[num_34 + num_842];
}
\end{lstlisting}
\end{minipage}%
\begin{tikzpicture}
  \draw[->] (0,0) -- (0.4,0);
\end{tikzpicture}%
\hspace{0.3cm}
\begin{minipage}{0.3\linewidth}
\begin{verbatim}
// Lexicalized tokens:
["int", "func", "_", "252", 
"()", "{", "int", "arr", "_"
... (tokens continue)
\end{verbatim}
\end{minipage}
\caption{Local Semantics Elimination (LSE) pipeline overview: Given a source code, the code turns into a semantic-less version using AST knowledge, and eventually, the lexicalized tokens are fed into \comp{}.}
\label{fig:orderoftokom}
\end{figure*}

\section{\hpcorpus{}: HPC Code Corpus}
\label{hpcor}

In order to build an HPC-specific model, we first decided to gather a dataset of HPC specific programs. Towards that end, we compiled \hpcorpus{}, a dataset of publicly-visible C and C++, and Fortran programs from GitHub~\cite{kadosh2023quantifying}. In this work we decided to focus on C and C++ languages only, and hence we used a subset of HPCorpus as shown in Table~\ref{tab:hpcorpus}.


In~\cite{kadosh2023quantifying}, we discovered that many of those repos employed one (or more) parallel API. For instance, 45\% employed shared memory parallelism with OpenMP (primarily for threading, but also for SIMD and GPU offloading), 27\% employed distribution with MPI, 21\% employed direct GPU programming (half with CUDA and half with OpenCL), and the rest mainly employed other threading APIs (such as TBB and Cilk).

\mypara{Training data preprocessing.} While many code LMs include natural language in their pre-training data, this will likely be unhelpful for HPC programmers and downstream HPC tasks of interest. On the other hand, it is much more critical that the model understands the structure of the code. To this end, we preprocess the code files in \hpcorpus{} such that no natural language is included; only structured blocks from deduplicated files are included; and code blocks are greater than 100 tokens and less than 1MB (as done in PolyCoder). By that, we gathered $\sim$155M functions that are more suitable for pre-training a code language model that emphasizes concise code structure.

\section{\comp{}: An HPC-specific code LM}
\label{CompCode}

To create a domain-specific model for HPC, we pre-trained a decoder-only transformer model on the language modeling objective (i.e., given past tokens as context, predict the next token in the code) using only C and C++ code from HPCorpus. We named this domain-specific model as \comp{}, in contrast to the multilingual Polycoder.

\mypara{Model size.} In the pursuit of deploying domain-specific small language models on average computer systems, careful consideration of model size becomes paramount. To ensure compatibility with such resource constraints, we aimed to design a model that strikes a balance between complexity and memory efficiency. The choice of a model comprising just under 1B parameters emerged as an optimal compromise. The decision is underpinned by the following calculations: Assuming each parameter is represented as a 32-bit floating-point number requiring 4 bytes, the formula \texttt{RAM size (in bytes)} = \texttt{Number of Parameters} $\times$ \texttt{Size of one parameter (in bytes)}. 
Consequently, for 0.9B parameters, the resulting RAM size amounted to 3.6GB. This selected model size not only accommodates a 4GB RAM constraint -- typically the maximum memory per core in HPC systems \cite{khan2021analysis} -- but also leaves additional headroom for other computational processes within the system to accommodate model inference. Note that this worst-case analysis did not consider inference-time optimizations such as quantization, or pruning, which would reduce memory consumption further.

We constructed \comp{} model of 0.9B parameters by reusing the PolyCoder 2.7B model architecture~\cite{xu2022systematic}, but reducing the number of layers. This led to an 8-layer model with a hidden dimension of 2560 and 32 attention heads per layer. We used the same tokenizer vocabulary (50,257 tokens). This resulted in an 889.3M parameter model ($\approx$ 0.9B parameters).

\mypara{Limitations.} Our domain-specific model inherits the principles of a left-to-right language model, which is especially amenable to code generation tasks. By adhering to the left-to-right nature, our model aligns with the established models such as CodeGPT (124M)~\cite{lu2021codexglue}, GPT-Neo~\cite{black2021gpt} and PolyCoder (2.7B)~\cite{xu2022systematic}, GPT-J (6B)~\cite{wang2021gpt}, Codex (12B)~\cite{chen2021evaluating}, StarCoder (15.5B)~\cite{li2023starcoder}, and GPT-NeoX (20B)~\cite{black2022gpt}, which are known for their proficiency in code completion.

However, we acknowledge the inherent challenge posed by the left-to-right approach, which limits the model's ability to consider context beyond the immediate token sequence. As part of our ongoing research, we are exploring strategies to address this limitation and further refine \comp{}'s capability to leverage broader contextual information, thereby advancing its effectiveness in the nuanced landscape of HPC code synthesis. 


\mypara{Pre-training details.} We pre-trained the \comp{} model using the GPTNeoX framework~\cite{andonian2023gpt-neox-library} on 4 NVIDIA A40 48GB GPUs with \texttt{fp16} precision. For pre-training, we used the Adam optimizer and followed a learning rate schedule with a linear warmup for the first 1\% of steps and a cosine decay over the remaining steps. Gradients were clipped at 1.0. Each training sample had a maximum length of 2048 tokens and was trained in mini-batches of 16 samples (4 per GPU). The model was trained for 160K steps at a learning rate of 0.00008. The training and validation losses initially start high at around 3.0, rapidly decrease over the first 25,000 steps, stabilize thereafter, converge closely without overfitting, and finally settle below 0.5 by approximately 150,000 steps, indicating minimal loss.


\begin{figure*}[!htbp]
\centering
    \begin{subfigure}{0.3\linewidth}
        \centering
        \begin{tikzpicture}
            \begin{axis}[
                ybar,
                width=6cm,
                height=5cm,
                ymax=25, 
                ytick={1, 5, 10, 15, 20, 25},
                ylabel={Size (B parameters)},
                xtick=data,
                nodes near coords,
                every node near coord/.append style={rotate=90, anchor=west, font=\small},
                xticklabel style={rotate=60},
                xticklabels={\comp{}, PolyCoder, GPT-Neo, GPT-J, Codex, StarCoder, GPT-NeoX},
                ymin=0,
                bar width=4pt, 
                label style={font=\small},
                tick label style={font=\small},
                grid style=dashed,
                ymajorgrids=true,
                ]

                \addplot[fill=archtBlue] coordinates {
                (1, 0.9) (2, 2.7) (3, 2.7) (4, 6) (5, 12) (6,15.5)  (7, 20)
                };
            \end{axis}
        \end{tikzpicture}
        \caption{Model Size\newline}
        \label{fig:model_size}
    \end{subfigure}%
    \hspace{0.5cm} 
    \begin{subfigure}{0.3\linewidth}
        \centering
        \begin{tikzpicture}
            \begin{axis}[
                ybar,
                width=6cm,
                height=5cm,
                ymax=50, 
                ytick={10,20,30,40,50},
                ylabel={Perplexity},
                xtick=data,
                nodes near coords,
                every node near coord/.append style={rotate=90, anchor=west, font=\small},
                xticklabel style={rotate=60},
                xticklabels={\comp{}, PolyCoder, GPT-Neo, GPT-J, Codex, StarCoder,  GPT-NeoX},
                ymin=0,
                bar width=4pt, 
                label style={font=\small},
                tick label style={font=\small},
                grid style=dashed,
                ymajorgrids=true,
                ]
                
                \addplot[fill=archtGreen] coordinates {
                (1, 3.51) (2, 2.33) (3, 3.69) (4, 2.82) (5, 2.55) (6,1.71)   (7, 2.37)
                };
                \addplot[fill=archtRed] coordinates {
                (1, 3.16) (2, 6.29) (3, 9.96) (4, 16.92) (5, 30.6) (6, 26.51) (7, 47.4)
                };
            \end{axis}
        \end{tikzpicture}
        \caption{Perplexity (in green) and Normalized\\-to-size Perplexity (in orange) for C}
        \label{fig:histogram_c}
    \end{subfigure}%
    \hspace{0.5cm} 
    \begin{subfigure}{0.3\linewidth}
        \centering
        \begin{tikzpicture}
            \begin{axis}[
                ybar,
                width=6cm,
                height=5cm,
                ymax=50, 
                ytick={10,20,30,40,50},
                ylabel={Perplexity},
                xtick=data,
                nodes near coords,
                every node near coord/.append style={rotate=90, anchor=west, font=\small},
                xticklabel style={rotate=60},
                xticklabels={\comp{}, PolyCoder, GPT-Neo, GPT-J, Codex, StarCoder, GPT-NeoX},
                ymin=0,
                bar width=4pt, 
                label style={font=\small},
                tick label style={font=\small},
                grid style=dashed,
                ymajorgrids=true,
                ]

                \addplot[fill=archtGreen] coordinates {
                (1, 3.69) (2, 2.99) (3, 2.87) (4, 2.47) (5, 1.95) (6, 2.01) (7, 2.32)
                };
                \addplot[fill=archtRed] coordinates {
                (1, 3.32) (2, 8.07) (3, 7.75) (4, 14.82) (5, 23.4) (6, 31.16) (7, 46.4)
                };
            \end{axis}
        \end{tikzpicture}
        \caption{Perplexity (in green) and Normalized\\-to-size Perplexity (in orange) for C++}
        \label{fig:histogram_cpp}
    \end{subfigure}
    \caption{Comparison of code language models based on their model size, perplexities, and normalized-to-size perplexities for C and C++. The results demonstrate that smaller models, such as \comp{}, tend to have much better normalized perplexity scores (lower is better), indicating better performance relative to their size. Data for PolyCoder, GPT-Neo, GPT-J, Codex, StarCoder, and GPT-NeoX are taken from \cite{xu2022systematic} and \cite{li2023starcoder}.}
    \label{fig:pplscores}
\end{figure*}

\section{Local Semantics Elimination (LSE)}  
\label{tokomp}


An effective code LM for HPC tasks must understand both the syntax and structure of the code, without memorizing potentially misleading human semantics~\cite{yang2023code} (e.g., variable names). We propose a preprocessing method based on an abstract syntax tree (AST) to remove misleading semantic information. A code LM that performs well on this anonymized preprocessed code is clearly relying on its understanding of code structure, rather than natural language-derived semantics. We name this preprocessing method as LSE (for local semantics elimination). LSE is inspired by input tokenization and more importantly ensures functionally correct and compilable code. 

Tokenizing code for LLMs necessitates specialized techniques to accommodate programming language syntax.\footnote{These approaches include utilizing BPE and subword tokenization akin to natural language~\cite{sennrich2015neural}, employing syntax-aware tokenization to identify language-specific elements like keywords and identifiers~\cite{zheng2022code}, constructing tokens based on the AST to capture structural information~\cite{xu2022survey}, implementing language-specific lexers following grammar rules~\cite{bui2023codetf}, preserving character integrity with character-level tokenization~\cite{kc2023neural}, tailoring tokenization to unique syntax rules, and leveraging dedicated code tokenization libraries.} LLMs geared towards code comprehension, such as GPT-3.5-Turbo for code,
likely combine several techniques, prioritizing syntax-aware tokenization to effectively process and generate code snippets in various programming languages and tasks. 
In short, LSE preprocessing (\autoref{fig:orderoftokom}) replaces variable names, numbers, and strings with random variable names and removes superfluous input (e.g., comments, extra whitespace).
The detailed steps are as follows:
\begin{enumerate}[wide, labelwidth=!, labelindent=0pt]
    \item \textit{AST Generation}: Parse the code using TreeSitter\footnote{\url{https://github.com/tree-sitter/tree-sitter}} or any suitable parser to generate an AST.
    \item \textit{Generate Replaced Code}: Create a version of the original code with anonymized variable names, numbers, and strings. The intuition behind this step is to eliminate misleading semantics, such as variable \texttt{i} being an index variable of a \texttt{for} loop in the C language.
    \item \textit{AST to Code}: Transform the updated AST back into code, while eliminating any comments that may interfere with anonymization. 
    \item \textit{Random Number Attachment}: For recurrent tokens (e.g., \texttt{var\_1} or \texttt{num\_2}), attach random integers from a predefined range (e.g., 1 to 1000) during tokenization. The attached numbers are randomly chosen without any relation to the type or order of the replaced tokens or the file/function length. This step also eliminates misleading semantics. For instance, if variable \texttt{i} is consistently replaced with \texttt{var\_1}, then the model may learn that \texttt{var\_1} is an index variable of \texttt{for} loops.
\end{enumerate}





\begin{figure*}[!htbp]
    \centering

    \begin{subfigure}[b]{0.2\textwidth}

   \begin{lstlisting}[breaklines=true, breakatwhitespace=true, basicstyle=\footnotesize\ttfamily, columns=fullflexible, numbers=none]
// INPUT to LLM:
\end{lstlisting} 

\begin{lstlisting}[breaklines=true, breakatwhitespace=true, basicstyle=\footnotesize\ttfamily, columns=fullflexible]
int main (int argc, char *argv[]) {
\end{lstlisting} 

\begin{lstlisting}[breaklines=true, breakatwhitespace=true, basicstyle=\footnotesize\ttfamily, columns=fullflexible, numbers=none]
// LLM COMPLETION:
\end{lstlisting} 

\begin{lstlisting}[breaklines=true, breakatwhitespace=true, basicstyle=\footnotesize\ttfamily, columns=fullflexible, firstnumber=2]
int i;
int n = argc - 1;
int m = argc - 2;
int k = argc - 3;
int l = argc - 4;
int r = argc - 5;
int c = argc - 6;
int d = argc - 7;
int e = argc - 8;
int f = argc - 9;
int g = argc - 10;
int h = argc - 11;
int i1 = argc - 12;
int i2 = argc - 13;
\end{lstlisting}        
\caption{Given function signature}
        \label{fig:subfig1}
    \end{subfigure}
    \hfill
    \begin{subfigure}[b]{0.38\textwidth}
       \begin{lstlisting}[breaklines=true, breakatwhitespace=true, basicstyle=\footnotesize\ttfamily, columns=fullflexible, numbers=none]
// INPUT to LLM:
\end{lstlisting} 
    \begin{lstlisting}[breaklines=true, breakatwhitespace=true, basicstyle=\footnotesize\ttfamily, columns=fullflexible]
int main (int argc, char *argv[]) {
int i,j;  
double x;  
double pi, sum = 0.0; 
double start, delta;  
step = 1.0/(double) steps;
for (j=1; j<= MAX_THREADS; j++) {
printf("running on %d threads: ", j);
omp_set_num_threads(j);  
    \end{lstlisting}
    \begin{lstlisting}[breaklines=true, breakatwhitespace=true, basicstyle=\footnotesize\ttfamily, columns=fullflexible, numbers=none]
// LLM COMPLETION:
\end{lstlisting} 
    \begin{lstlisting}[breaklines=true, breakatwhitespace=true, basicstyle=\footnotesize\ttfamily, columns=fullflexible, firstnumber=10]
start = omp_get_wtime();
for (i=0; i<steps; i++) {    
x = (double) i;
pi = 3.141592653589793238
4626433832795028841971693
9937510582097494459230781
6406286208998628034825342...
    \end{lstlisting}
        \caption{Given initialization}
        \label{fig:subfig2}
    \end{subfigure}
    \hfill
    \begin{subfigure}[b]{0.38\textwidth}
       \begin{lstlisting}[breaklines=true, breakatwhitespace=true, basicstyle=\footnotesize\ttfamily, columns=fullflexible, numbers=none]
// INPUT to LLM:
\end{lstlisting} 
    \begin{lstlisting}[breaklines=true, breakatwhitespace=true, basicstyle=\footnotesize\ttfamily, columns=fullflexible]
int main (int argc, char *argv[]) {
int i,j;
double x;
double pi, sum = 0.0;
double start, delta;
step = 1.0/(double) steps;
for (j=1; j<= MAX_THREADS; j++) { 
printf("running on %d threads: ", j);
omp_set_num_threads(j);  
sum = 0.0;
double start = omp_get_wtime();  
#pragma omp parallel for reduction(+:sum) private(x) 
    \end{lstlisting}
    \begin{lstlisting}[breaklines=true, breakatwhitespace=true, basicstyle=\footnotesize\ttfamily, columns=fullflexible, numbers=none]
// LLM COMPLETION:
\end{lstlisting} 
        \begin{lstlisting}[breaklines=true, breakatwhitespace=true, basicstyle=\footnotesize\ttfamily, columns=fullflexible, firstnumber=13]
for (i=0; i<steps; i++) {
    x = (double) i;
    sum += x;} ...
    \end{lstlisting}
        \caption{Given context of parallel execution}
        \label{fig:subfig3}
    \end{subfigure}
    \caption{Evaluating code completion performance of HPC code by the foundation models (PolyCoder example): Evaluating machine-generated code, given different contexts of the initial HPC codes and measuring the similarity to the true reference.}
    \label{fig:bleutests}
\end{figure*}

\begin{figure*}
    \centering
    
\begin{subfigure}{0.45\linewidth}
        \begin{tikzpicture}
            \begin{axis}[
                legend columns=2,
                ybar,
                bar width=0.15cm, 
                ylabel={CodeBLEU},
                symbolic x coords={context-100, context-300, context-600},
                xtick=data,
                ymin=0,
                ymax=1.5,
                enlarge x limits=0.2,
                nodes near coords,
                every node near coord/.append style={rotate=90, anchor=west, font=\scriptsize},
                ybar=3pt, 
                legend style={at={(0.5,-0.3)},anchor=north,legend columns=-1},
                grid style=dashed,
                height=4cm,
                width=\linewidth,
        ymajorgrids=true,
            ]
            \addplot[fill=archtBlue] coordinates {
                (context-100, 0.617)
                (context-300, 0.750)
                (context-600, 0.846)
            };
            \addplot[fill=lightBlue] coordinates {
                (context-100, 0.642)
                (context-300, 0.722)
                (context-600, 0.847)
            };
            \addplot[fill=archtRed] coordinates {
                (context-100, 0.451)
                (context-300, 0.635)
                (context-600, 0.781)
            };
            \addplot[fill=lightRed] coordinates {
                (context-100, 0.387)
                (context-300, 0.595)
                (context-600, 0.743)
            };
            \addplot[fill=archtGreen] coordinates {
                (context-100, 0.437)
                (context-300, 0.579)
                (context-600, 0.585)
            };
            \addplot[fill=lightGreen] coordinates {
                (context-100, 0.525)
                (context-300, 0.582)
                (context-600, 0.602)
            };

            \legend{\comp{}, \comp{} + LSE, PolyCoder, PolyCoder + LSE, GPT-3.5, GPT-3.5 + LSE}
            \end{axis}
        \end{tikzpicture}
        \caption{Code completion performance on \textit{General} dataset.}
        \label{fig:hpc_code_bleu}
    \end{subfigure}
        \begin{subfigure}{0.45\linewidth}
        \begin{tikzpicture}
            \begin{axis}[
                legend columns=2,
                ybar,
                bar width=0.15cm, 
                ylabel={CodeBLEU},
                symbolic x coords={context-100, context-300, context-600},
                xtick=data,
                nodes near coords,
                every node near coord/.append style={rotate=90, anchor=west, font=\scriptsize},
                ybar=3pt, 
                ymin=0,
                ymax=1.5,
                height=4cm,
                width=\linewidth,
                enlarge x limits=0.2,
                legend style={at={(0.5,-0.3)},anchor=north,legend columns=-1},
                grid style=dashed,
        ymajorgrids=true,
            ]
            \addplot[fill=archtBlue] coordinates {
                (context-100, 0.636)
                (context-300, 0.765)
                (context-600, 0.841)
            };
            \addplot[fill=lightBlue] coordinates {
                (context-100, 0.638)
                (context-300, 0.768)
                (context-600, 0.856)
            };
            \addplot[fill=archtRed] coordinates {
                (context-100, 0.605)
                (context-300, 0.742)
                (context-600, 0.838)
            };
            \addplot[fill=lightRed] coordinates {
                (context-100, 0.374)
                (context-300, 0.585)
                (context-600, 0.735)
            };
            \addplot[fill=archtGreen] coordinates {
                (context-100, 0.429)
                (context-300, 0.566)
                (context-600, 0.561)
            };
            \addplot[fill=lightGreen] coordinates {
                (context-100, 0.517)
                (context-300, 0.572)
                (context-600, 0.595)
            };

            \legend{\comp{}, \comp{} + LSE, PolyCoder, PolyCoder + LSE, GPT-3.5, GPT-3.5 + LSE}
            \end{axis}
        \end{tikzpicture}
        \caption{Code completion performance on \textit{OpenMP} dataset.}
        \label{fig:omp_code_bleu}
    \end{subfigure}%
    \caption{Code Completion Performance on General and OpenMP Datasets –- CodeBLEU scores (higher is better) for \comp{}, PolyCoder, and GPT-3.5 models, both with and without Local Semantic Elimination (LSE), across varying context lengths (100, 300, and 600 tokens). \comp{} and \comp{} + LSE consistently outperform other models, with the addition of LSE generally enhancing performance across all models.} 
    \label{fig:code_bleu}
\end{figure*}

\section{\comp{} Evaluation}
\label{section:intrinsic_eval}
\subsection{Language Modeling Evaluation}

We start by measuring the perplexity of \comp{} and other LLMs on unseen test programs in C and C++.

\mypara{Metric: Perplexity.} 
Perplexity is a metric commonly employed in natural language processing and language modeling to assess the efficacy of a probabilistic model in predicting a given sequence of tokens. It serves as a measure of the model's uncertainty or confusion when assigning probabilities to a set of observed data. A lower perplexity value signifies the better predictive performance of a model. In the context of our research, we measured perplexity to evaluate the predictive capabilities of the pre-trained \comp{} language model. That is, this assessed how well the model assigned probabilities to the token sequences in the C and C++ portions of the \hpcorpus{} test set.

\mypara{Setup.} For this test, we use the training procedure outlined earlier to pre-train \comp{} on \hpcorpus{}. We then measure perplexity on C and C++ code from the test set of \hpcorpus{}. We then compared this perplexity to the perplexity scores of various pre-trained LLMs based on the published works~\cite{xu2022systematic, li2023starcoder}. The StarCoder values are with a 2K context window, similar to \comp{}'s 2048 context window.

\mypara{Results.} The perplexity comparison can be seen in \autoref{fig:pplscores}. We can see that despite the much smaller size of \comp{}, it suffers from minor performance degradation than much larger models. For C language, it is performing better than the 3x-larger GPT-Neo model.



\subsection{Code Completion Evaluation}

In this second evaluation, we assess the ability of \comp{} and other LLMs in completing a block of code when provided with varying amounts of prior context. While perplexity provides information about the uncertainty of the model, it does not necessarily measure the quality of the generated code. For this reason, we further evaluate code understanding with the CodeBLEU~\cite{ren2020codebleu} score.


\mypara{Metric: CodeBLEU score.} CodeBLEU~\cite{ren2020codebleu} is the metric of choice in the code completion tasks. CodeBLEU amalgamates the robustness of BLEU score~\cite{papineniBleuMethodAutomatic2002}(from NLP), incorporating n-gram matching, with an innovative integration of code syntax and semantics through AST and data-flow structures. This holistic approach provides a nuanced evaluation that extends beyond token matching, considering the significance of keywords, syntactic accuracy, and semantic correctness. In the case of HPC code without natural language, CodeBLEU is the appropriate metric to check similarity to generations.

\mypara{Setup.} For this evaluation, we first devise two sub-datasets: (1) \textit{General} dataset of about 20k examples of C and C++ programs from \hpcorpus{} with HPC-orientation\footnote{Ones that contained OpenMP's \texttt{parallel for} pragmas. However, those pragmas were intentionally removed to keep clean C and C++ programs that have HPC orientation} and (2) \textit{OpenMP} dataset of about 20k examples of C and C++ programs containing OpenMP code (e.g., \texttt{parallel for)}. The method to test the model's understanding is by incrementally supplying it more parts of those programs (first 100, 300, and 600 tokens, out of an average of 1200 tokens per code) and comparing the code generated by the model with the expected ground-truth code using CodeBLEU. Demonstration of the idea is presented in \autoref{fig:bleutests}.

\mypara{Results.} \autoref{fig:hpc_code_bleu} shows code completion performance of the models on \textit{General} dataset. For a context of the first 100 tokens, \comp{} achieves a CodeBLEU score of 0.62, while PolyCoder and GPT-3.5 attain scores of 0.45 and 0.44, respectively. This discrepancy can be attributed to PolyCoder's diverse training across programming languages and GPT-3.5's general nature, resulting in versatile but less HPC-specific answers. Expanding the context window to 600 tokens further widens the CodeBLEU score gap between \comp{} and GPT-3.5, emphasizing lack of HPC knowledge in GPT. Overall, the model's reduced orientation towards HPC codes accentuates the discernible gap in HPC code knowledge among the models.

Similar trends are evident on \textit{OpenMP} dataset in \autoref{fig:omp_code_bleu}, depicting the CodeBLEU scores of models when completing functions containing OpenMP pragmas. Generally, the results are slightly lower than those on \textit{General} dataset, underscoring the models' limited understanding of OpenMP.

It is noteworthy that, despite the LSE pre-processing having a minor impact on the performance of \comp{} and GPT-3.5, this code representation significantly impairs the performance of PolyCoder. This observation suggests that PolyCoder heavily relies on local semantics for code completion.

\section{Related Work}
\label{related}

We present the related work along two directions: 1) code LMs that are not specifically designed for HPC tasks but can solve HPC tasks, 2) LLMs designed specifically for HPC tasks.

The code LM literature is somewhat divided into NLP-oriented approaches and software engineering approaches. Popular LLMs such as GPT-3.5, GPT-4, LLaMa, or even code specific LLMs such as CodeT5~\cite{wang2021codet5identifierawareunifiedpretrained}, etc., are trained on code datasets that also contain natural language comments. As a result, most of these models, if not all, can solve programming tasks with or without natural language prompts. These NLP-oriented approaches typically evaluate code LMs with perplexity and datasets like HumanEval~\cite{chen2021evaluating} or Mostly Basic Programming Problems~\cite{austinProgramSynthesisLarge2021}, which assess both natural language comprehension and general reasoning ability of a model along with code comprehension. These models also solve coding-specific tasks such as code search, code completion, etc. In addition to natural language and programming-specific tasks, these models can also solve HPC tasks to certain extent. Because the training datasets of these models most likely contain HPC code also (in APIs such as OpenMP, CUDA, etc.), these models possess some understanding of HPC languages also. Our experiments with GPT-3.5 revealed that it had a reasonable understanding of code parallelization problem (e.g., analyzing if a loop can be parallelized). However, our experiments also revealed that these LLMs have limited understanding of HPC tasks --- the same observation also reported by Nichols et al. in their evaluation of LLMs for HPC tasks~\cite{nicholsCanLargeLanguage2024}. Specifically, their experimental results reveal several limitations of existing LLMs for HPC tasks.

Given the limitations of popular LLMs on HPC tasks and also their expensive training costs, several works have recently explored HPC specific LMs. Specifically, several groups have developed LMs for popular HPC problems such as OpenMP pragma prediction and generation~\cite{chen2023lm4hpc, chen2024ompgpt, harel2023learning, kadosh2023advising, kadosh2023pragformer, mahmud2023autoparllm, nichols2023modeling, chen2023learning}, MPI code generation~\cite{schneider2023mpi, schneider2024mpirigen}, and race detection~\cite{ding2023hpc}. 
We found that these works mostly finetune a pre-trained LM, which are larger in size than required for their HPC tasks. We believe that our work complements these papers by demonstrating a smaller, domain-specific LM that can solve HPC tasks.


\section{Conclusion \& Future Work}
\label{conclusion}


Existing LLMs, such as GPT-3.5, are multi-lingual and are trained on languages unrelated to HPC. This phenomenon leads to huge model sizes and demands expensive compute resources to train. Thus, we decided to evaluate if we can build a smaller, domain-specific model that can perform similar, if not better, than existing LLMs on HPC-related programming tasks. Towards that end, we built \comp{} using an existing code-oriented LLM but by carefully selecting the model size such that the resulting model can fit on a commodity hardware. Our experimental results demonstrate that \comp{}, although orders of magnitude smaller in size than existing LLMs, performs similar to the existing LLMs in terms of language comprehension task (perplexity score), while outperforming them in code completion task (CodeBLEU score) for general-purpose and HPC-specific C and C++ programs. 

In the near future, we intend to integrate additional code representations, such as the data-flow graph (DFG) and the intermediate representation (IR)~\cite{grossman2023compile}, to enhance model understanding as shown in the closely related works~\cite{guo2020graphcodebert, szafraniec2022code}. Moreover, we believe that pre-training on a compilable subset of \hpcorpus{} will enhance the performance of the model for compilation-oriented tasks (as partially demonstrated in~\cite{chen2023compcodevet}). Then, we intend to fine-tune those pre-trained models for HPC-specific downstream tasks, such as OpenMP pragma generation~\cite{chen2024ompgpt, kadosh2023advising, nichols2023modeling} and MPI domain decomposition distribution~\cite{nichols2023modeling, schneider2023mpi, schneider2024mpirigen}. In general, our research vision is to systematically analyze each and every element of existing LLMs (model architecture, dataset, etc.) and redesign them as needed for HPC-specific tasks.


\bibliographystyle{IEEEtran}
\bibliography{IEEEabrv, references} 

\end{document}